\documentclass[cits]{PoS} 
\title{Pseudoscalar-Meson Form Factors:\\A Fresh Look by QCD Sum
Rules}\ShortTitle{Pseudoscalar-Meson Form Factors: A Fresh Look by
QCD Sum Rules}
\author{Irina Balakireva\\D.~V.~Skobeltsyn Institute of Nuclear
Physics, Moscow State University, 119991, Moscow, Russia\\E-mail:
\email{iraxff@mail.ru}}
\author{\speaker{Wolfgang Lucha}\\Institute for High Energy
Physics, Austrian Academy of Sciences, Nikolsdorfergasse 18,
A-1050 Vienna, Austria\\E-mail: \email{Wolfgang.Lucha@oeaw.ac.at}}
\author{Dmitri Melikhov\\Institute for High Energy Physics,
Austrian Academy of Sciences, Nikolsdorfergasse 18, A-1050 Vienna,
Austria,\\Faculty of Physics, University of Vienna, Boltzmanngasse
5, A-1090 Vienna, Austria, and\\D.~V.~Skobeltsyn Institute of
Nuclear Physics, Moscow State University, 119991, Moscow,
Russia\\E-mail: \email{dmitri\_melikhov@gmx.de}}

\abstract{Confronted with some surprising claims about the either
experimentally measured or theoretically expected dependences on
the involved momentum transfer of various form factors of
pseudoscalar mesons, we reassess the present status of these
objects by means of QCD sum rules. This approach provides
well-developed and very efficient tools to relate in an analytical
manner the parameters of {\em quantum chromodynamics\/}
(QCD)---the quantum field theory that describes the strong
interactions responsible for the formation of hadronic bound
states---to the empirical features of such particles: Matrix
elements of appropriately chosen {\em products of interpolating
currents\/} that carry the quantum numbers of the hadrons under
study are evaluated at both hadron and QCD level. In the latter
case, all these nonlocal operators are expressed as series of
local operators by Wilson's operator product expansion, with
coefficients determined from perturbation theory. For vacuum
expectation values, this introduces universal vacuum condensates
that parameterize the nonperturbative contributions. Our ignorance
about the higher hadron states is masked by quark--hadron duality
assuming mutual cancellations of the contributions of hadronic
excitations and continuum and of {\em perturbative\/} QCD beyond
certain effective thresholds. Within this framework we show that a
few theoretical findings for the charged-pion elastic form factor
and one experimental result for the neutral-pion-to-photon
transition form factor are at odds with very general, and likely
sound, fundamental considerations.}

\FullConference{Xth Quark Confinement and the Hadron
Spectrum\\8--12 October 2012\\TUM Campus Garching, Munich,
Germany}

\begin{document}\section{Dispersive Local-Duality QCD Sum Rules for
Pseudoscalar-Meson Form Factors}
\begin{figure}[h]\begin{center}\begin{tabular}{c}
\includegraphics[scale=.52127]{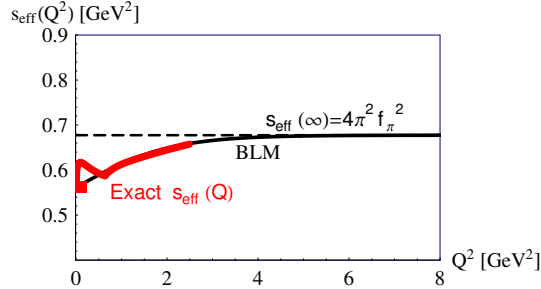}\end{tabular}
\caption{Naive LD modelling (BLM) \cite{BLM} of the {\em exact\/}
effective threshold for $F_\pi(Q^2)$ fixed by
experiment~\cite{Ee}.}\label{Fig:s}\end{center}\end{figure}

We analyze the dependences on the
involved momentum transfer squared $Q^2$ of the elastic form
factor $F_\pi(Q^2)$ of the charged pion and of the form factor
$F_{\pi\gamma}(Q^2)$ describing the transition
$\pi^0\leftrightarrow\gamma\,\gamma^*$ of the neutral pion to a
real and a virtual photon by dispersive QCD sum rules deduced from
vacuum expectation values of products of the three adequate
interpolating or electromagnetic currents \cite{BLM,LM}. In the
limit of local duality (LD) \cite{LD} the sum rules relate the
form factors to dispersion integrals over perturbatively deducible
spectral densities. The $Q^2$-dependent upper integration limits,
the effective thresholds $s_{\rm eff}(Q^2),$ then encode the
nonperturbative effects. In terms of strong coupling $\alpha_{\rm
s}(Q^2)$ and $f_\pi=130\;\mbox{MeV},$ for $Q^2\to\infty$
factorization implies $Q^2\,F_\pi(Q^2)\to8\pi\,\alpha_{\rm
s}(Q^2)\,f_\pi^2,$ $Q^2\,F_{\pi\gamma}(Q^2)\to\sqrt{2}\,f_\pi,$
and thus $s_{\rm eff}(\infty)=4\pi^2\,f_\pi^2$ \cite{F}. Even if
supported by quantum-mechanical solutions \cite{LS}, modelling of
$s_{\rm eff}(Q^2<\infty)$ is non-trivial \cite{LMS}. To sharpen
our arguments, we define equivalent effective thresholds: sum
rules with such integration limit reproduce experiment or any
theoretical result exactly (Fig.~\ref{Fig:s}).

\vspace{-3.50644ex}\section{Charged-Pion Elastic and
Neutral-Pion-to-Photon Transition Form Factors \cite{BLM,LM}}
\begin{figure}[h]\begin{center}\begin{tabular}{c}
\includegraphics[scale=.52127]{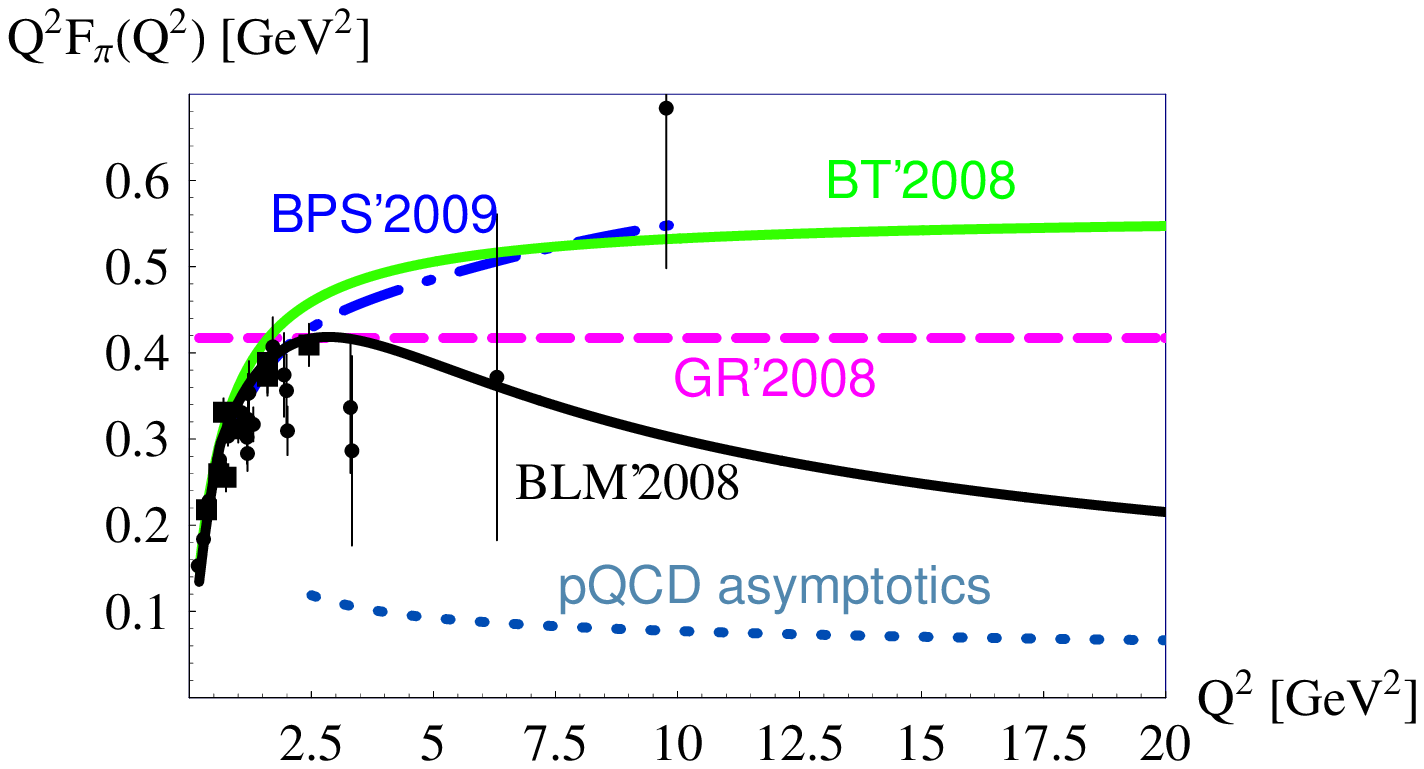}
\includegraphics[scale=.52127]{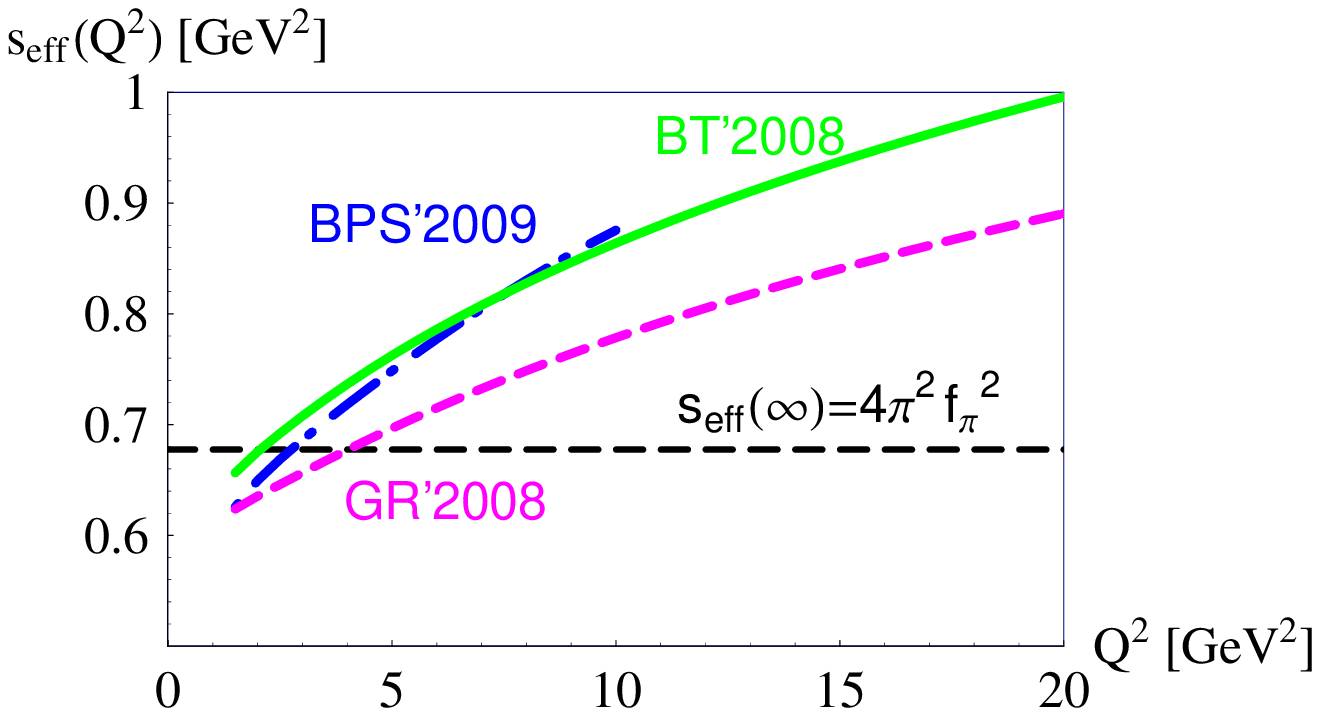}\end{tabular}
\caption{The $Q^2$ behaviour of the elastic pion form factor
$F_\pi(Q^2)$ obtained \cite{T} by Brodsky and de T\'eramond
(BT'2008), Grigoryan and Radyushkin (GR'2008), and Bakulev,
Pimikov, and Stefanis (BPS'2009)---and of the equivalent effective
thresholds $s_{\rm eff}(Q^2)$---is in clear conflict with {\em LD
model\/} expectations \cite{BLM} (BLM'2008).}\label{Fig:e}
\end{center}\end{figure}

Our
model interpolates between the large-$Q^2$ asymptote and the
empirical low-$Q^2$ behaviour of $s_{\rm eff}(Q^2)$ \cite{BLM}
(Fig.~\ref{Fig:s}). Not {\em all\/} other approaches \cite{T}
comply with the resulting form of $F_\pi(Q^2)$~(Fig.~\ref{Fig:e}).

\vspace{-.5ex}

\begin{figure}[h]\begin{center}\begin{tabular}{c}
\includegraphics[scale=.5232]{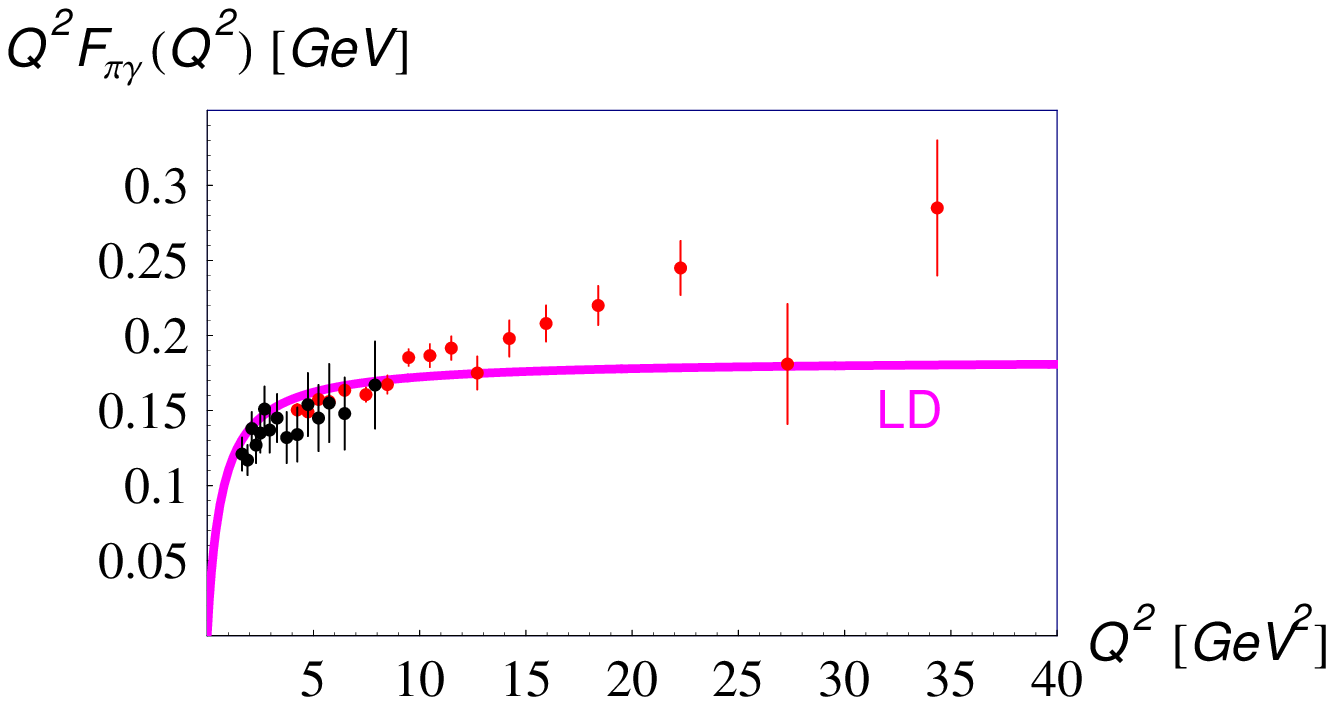}
\includegraphics[scale=.5232]{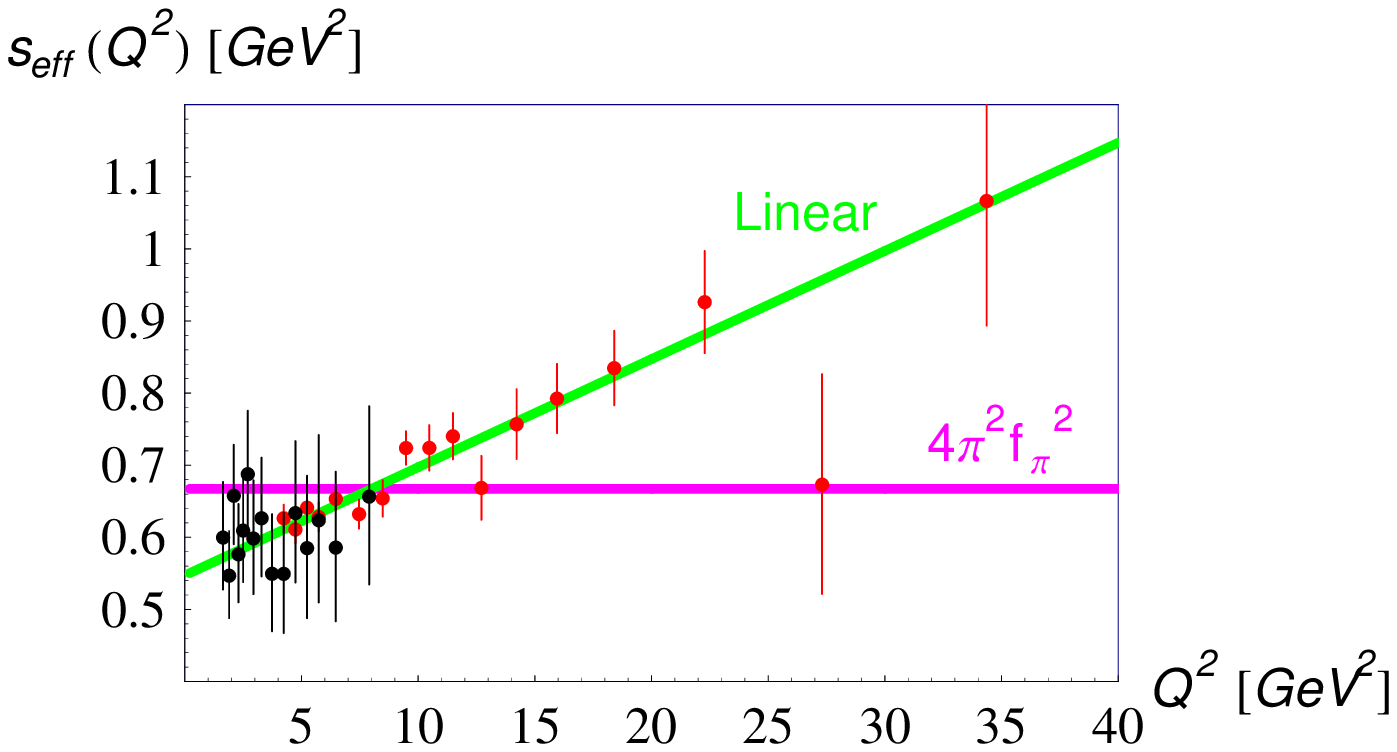}\end{tabular}
\caption{The {\sc BaBar} results for $F_{\pi\gamma}(Q^2)$
\cite{Et} (red dots), unlike those by CELLO and CLEO (black
dots)~\cite{Et}, are not compatible with LD sum-rule predictions
as their incorporation would require a linear rise of~$s_{\rm
eff}(Q^2).$}\label{Fig:t}\end{center}\end{figure}

Regarding pseudoscalar-meson transition form factors, LD sum rules
perform satisfactorily for
$(\eta,\eta',\eta_c)\leftrightarrow\gamma\,\gamma^*$ but do not
reproduce a {\sc BaBar} claim \cite{Et} of
$Q^2\,F_{\pi\gamma}(Q^2)$ rising with $Q^2$ beyond~the {\em LD
asymptote\/} $\sqrt{2}\,f_\pi$ \cite{BLM,LM} (Fig.~\ref{Fig:t});
confident in our approach, we feel that this mismatch casts some
doubt on the {\sc BaBar} measurement. Recent Belle observations
\cite{Et} lend support to our point~of view.

\noindent{\bf Acknowledgments.} D.M.\ was supported by the
Austrian Science Fund (FWF), project no.~P22843.

\end{document}